\documentclass[namedreferences]{solarphysics}

\usepackage[hyperref,optionalrh,showbiblabels]{spr-sola-addons} 
\usepackage{hyperref}
\hypersetup{
    colorlinks=true,
    linkcolor=red,
    filecolor=magenta,      
    urlcolor=cyan,
}

\usepackage{graphicx}        
\usepackage{color}           
\usepackage{breakurl}        

\newcommand{\aap}{    {\it Astron. Astrophys.}}

\newcommand{\apj}{    {\it Astrophys. J.}}
\newcommand{\apjl}{   {\it Astrophys. J. Lett.}}

\newcommand{\grl}{    {\it Geophys. Res. Lett.}}

\newcommand{\solphys}{{\it Solar Phys.}}
 
\newcommand{\ssr}{    {\it Space Sci. Rev.}} 
\chardef\us=`\_

\begin{document}

\begin{article}
\begin{opening}

\title{Radio Interferometric Observations of the Sun Using Commercial Dish TV Antennas}

\author[addressref={aff1},email={gireesh@iiap.res.in}]{\inits{G. V. S. Gireesh}
\fnm{G. V. S. Gireesh}}
\author[addressref={aff1},email={kathir@iiap.res.in}]{\inits{C. Kathiravan}
\fnm{C. Kathiravan}}
\author[addressref={aff1},email={indrajit@iiap.res.in}]{\inits{Indrajit V. Barve}
\fnm{Indrajit V. Barve}}
\author[addressref={aff1},email={ramesh@iiap.res.in}]{\inits{R. Ramesh}
\fnm{R. Ramesh}\orcid{0000-0003-2651-0204}}
\address[id=aff1]{Indian Institute of Astrophysics, 2nd Block, Koramangala, Bangalore 560034}

\runningauthor{G. V. S. Gireesh et al.}
\runningtitle{Radio Observations Using Commercial Dish TV Antennas}

\begin{abstract}
The radio astronomy group in the Indian Institute of Astrophysics (IIA) has been carrying out routine observations of radio emission from the solar corona at low frequencies (${\approx}$40-440MHz) at the Gauribidanur observatory, about 100km north of Bangalore. 
Since IIA has been performing regular observations of the solar photosphere and chromosphere using different optical telescopes in its Kodaikanal Solar Observatory (KSO) also\footnote{\textit{see}~\url{https://www.iiap.res.in/kodai.htm}}, the possibilities of obtaining two-dimensional radio images of the solar chromosphere using low-cost instrumentation to supplement the optical observations are being explored.
As a part of the exercise, recently the group had developed prototype
instrumentation for interferometric observations of radio emission from the solar chromosphere at high frequencies (${\approx}$11.2GHz) using two commercial dish TV antennas. The hardware set-up and initial observations
are presented. 
\end{abstract}
\keywords{Sun, radio emission. Sun, eclipse. Radio telescopes, interferometry.}
\end{opening}

\section{Introduction}
\label{S-Intro} 

Commercial dish TV antennas are parabolic structures designed to receive radio waves from a communication satellite. The antennas and the associated front end receiver systems have improved with advances in the TV systems. They operate typically over the frequency range 
10.7-11.7GHz (Ku-band) and provide very good signal-to-noise ratio (SNR). It is well known that the Sun emits intense radio emission in the above frequency range with brightness temperature $T_{b}{\sim}10^{4}$K. The emission is primarily due to thermal free-free mechanism, and originates in the upper chromosphere. The observed $T_{b}{=}T_{e}(1-e^{-\tau})$, where $T_{e}$ is the electron temperture and $\tau$ is the optical depth of the medium. Therefore, estimates of $T_{b}$ allows direct measurement of $T_{e}$, a `true' temperature since it is independent of abundances or atomic physics.
The optical depth ($\tau$) factor in the above equation can be addressed using simultaneous observations at more than one frequency (see for e.g. \citealp{Zirin91}). It is well established that $T_{b}$ is one of the basic properties of the Sun.
Hence its routine monitoring is widely used as a diagnostic tool to understand the different types of temporal changes 
in the solar emission \citep{Leblanc69,Tanaka73,Lantos75,Kruger86,Ramesh00,Ramesh01,Ramesh06b,Kathiravan02,
Kathiravan07,
Selhorst11,Ramesh12b,Saint13,Shibasaki13,Tapping13,Sasikumar14,Mercier15,Ramesh20,
Anshu21}. 
Motivated by the low-cost e-CALLISTO spectrometers which are successfully used for solar observations in the frequency range ${\sim}45{-}870$MHz \citep{Benz09} and 
considering off the shelf availability of dish TV antennas as well its associated receiver components at relatively low prices, we attempted observations of the Sun in the microwave frequency range mentioned above with a pair of commercial dish TV antennas. 
The observations were made in the correlation interferometer mode since it provides better sensitivity. Further, any contribution from the galactic background will also be negligible \citep{Kraus86}. A two-element radio interferometer is also the building block of a synthesis imaging array \citep{Thompson04}. 
Some of the correlation radio interferometer arrays that routinely observe the Sun in the microwave frequency range with customized antennas are Nobeyama RadioHeliograph (NoRH; \citealp{Nakajima94}), Siberian Solar Radio Telescope (SSRT; \citealp{Grechnev03}), Expanded Owens Valley Solar Array (EOVSA; \citealp{Nita16}), and MingantU SpEctral Radioheliograph (MUSER; 
\citealp{Yan21}).  
  
\section{Antenna and Receiver System} 
\label{S-Ars}  
      
\begin{figure} 
\centerline{\includegraphics[width=\textwidth,clip=]{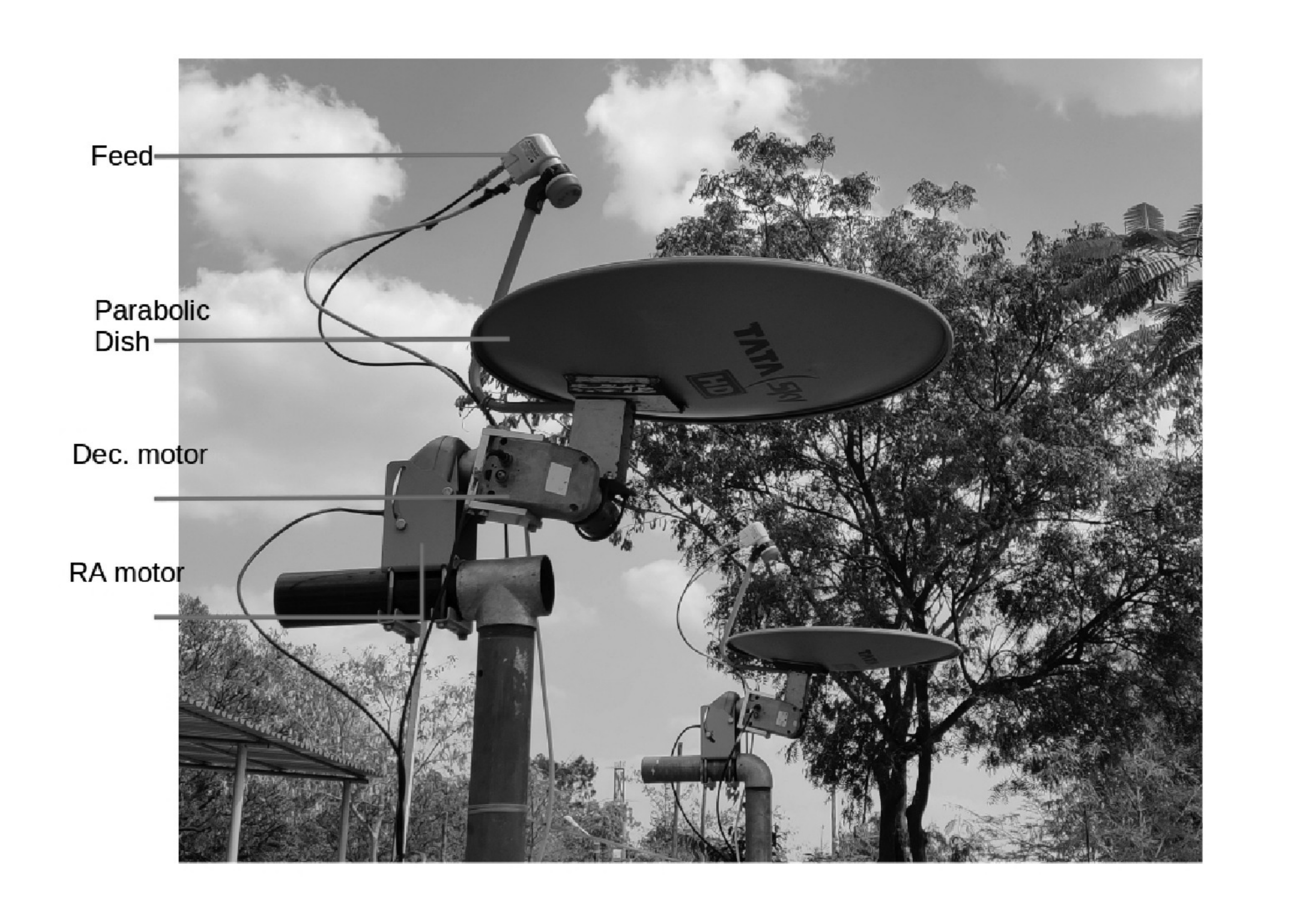}}
\caption{The commercial dish TV antennas used in the Gauribidanur observatory for solar observations in the correlation interferometer mode.}
\label{figone}
\end{figure}

\begin{figure} 
\centerline{\includegraphics[width=\textwidth,clip=]{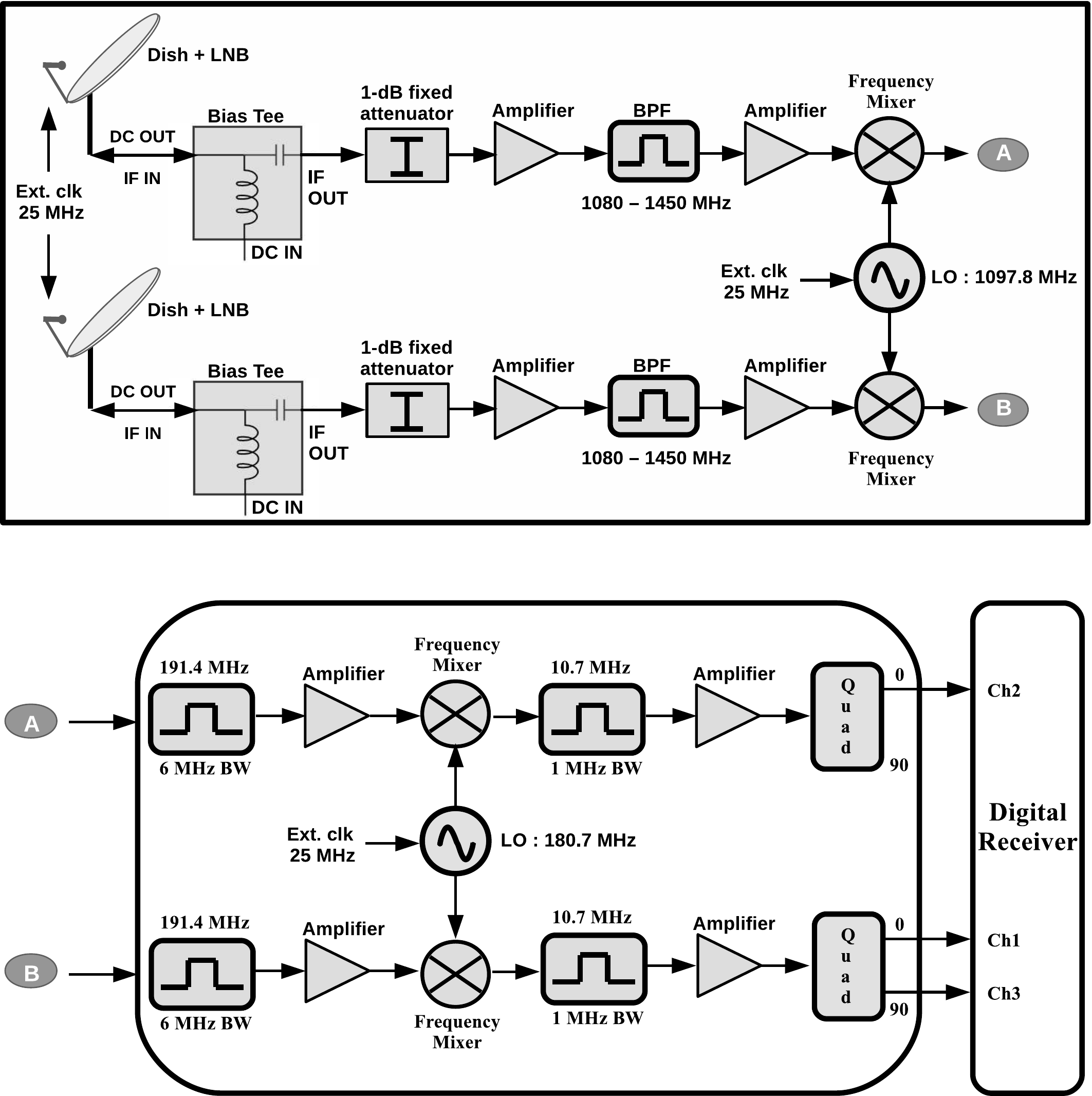}}
\caption{The analog receiver section corresponding to the two-element interferometer in Figure~\ref{figone}. `Quad' is quadature power-splitter. The 25MHz external clock is a common clock.}
\label{figtwo}
\end{figure}

Figure~\ref{figone} shows the two commercial dish TV antennas set-up in the Gauribidanur observatory\footnote{\textit{see}~\url{https://www.iiap.res.in/?q=centers/radio}} for solar observations \citep{Sastry95,Ramesh11}. Each dish is parabolic reflector of diameter {D$\approx$}62cm. 
Its an equatorial mount system.
There is a feed horn and Low Noise Block (LNB, otherwise called the front end receiver) at the focus of the reflector. They receive and amplify the signal reflected by the parabolic dish. 
The LNB\footnote{\textit{see}~\url{https://www.solid.sale/lnbf/ku-band-lnb/fs-108-ku-band-lnb}}
has a noise figure (NF)${\approx}$0.6dB (corresponding noise temperature is ${\approx}$45K), and gain (G)${\approx}$60dB. The characteristic impedance ($\rm Z_{o}$) of the LNB is ${\approx}75{\Omega}$. But the co-axial cables and other components used in the subsequent stages of the analog receiver system 
have $\rm Z_{o}{\approx}50{\Omega}$. Due to this, there will be reflection of signal from the load and hence standing waves in the corresponding signal path. 
But we found that the reflected power due to the above impedance mismatch is small (${\approx}$4\%). The measured reflection coefficient ($\Gamma$) is ${\approx}$0.2. This indicates that the amount of transmitted power that gets attenuated due to signal reflection is, 10log(1-${\Gamma}^2)\,{\approx}$\,0.2dB. Nevertheless we have connected a 1dB fixed attenuator between the LNB and the following amplifier (see Figure~\ref{figtwo}) to minimize the effects of the reflected 
signal\footnote{\textit{see}~\url{https://www.minicircuits.com/app/AN70-001.pdf}, \textit{etc.}}.
The response pattern (`beam') of each dish has a theoretical half-power beamwidth 
(HPBW${\approx}70^{\circ}\lambda$/D) of ${\approx}3.02^{\circ}{\times}3.02^{\circ}$ (R.A.${\times}$Dec.) at a typical frequency like 11.2GHz. Note that the above formula assumes that the electric field distribution is uniform across the aperture (see for e.g. \citealp{Kraus86}). The feed horn receives radio frequency (RF) signal in the frequency range ${\approx}$10.7-11.7GHz. The LNB has a frequency translating unit or a `mixer' which converts the aforementioned RF signal to an intermediate frequency (IF) range of ${\approx}$950-1950MHz by using a local oscillator (LO) signal of frequency 9.75GHz.

We had set-up a correlation interferometer with a separation of ${\approx}$2.5m between the two dish antennas. The baseline between the antennas is oriented in the East-West direction. So, the theoretical angular resolution in that direction (for observations near the zenith), specified by separation between the interference fringes, is 
${\approx}37^{\prime}$ at 11.2GHz. The angular size of the Sun at 11GHz 
is ${\approx}33^{\prime}$ \citep{Furst79}. Since this is smaller than the aforementioned fringe spacing, Sun can be assumed to be a `point' source for our observations. The corresponding resolution in the North-South direction is 
${\approx}3.02^{\circ}$, i.e. the HPBW of the dish mentioned above.
A pair of motors, one for R.A. and other for Dec., are used for each dish antenna to tilt it towards the source position in the sky. Both the motors are controlled by a common Arduino board interfaced to a computer\footnote{\textit{see}~\url{http://www.e-callisto.org/Hardware/Callisto-Hardware.html}}. In the interferometric mode of observations the phase of LO signal at both the feeds must be synchronized. To achieve temporal coherence, we used a common external 25MHz clock signal to trigger the oscillators in the two LNBs. DC power supply for each LNB was provided using a 
Bias-Tee. 
It is a 3-port network\footnote{\textit{see}~\url{https://www.microwaves101.com/encyclopedias/bias-tee}} having (i) RF port where only RF signal can be extracted by blocking DC with the help of a capacitor, (ii) RF+DC port where DC is sent to bias the module (LNB in our case) and receive RF signal from the latter, and (iii) DC port where the DC source for biasing the module is connected. We used `Nikou' make (10MHz-6GHz) Broadband Radio Frequency Microwave Coaxial Bias-Tee 1-50V, 0.5A (max)
for the present work. The IF signal from each of the two antennas are independently amplified, filtered 
(${\approx}$1080-1450MHz) and then transmitted to an analog receiver after downconverting to 
${\approx}$191.4MHz by mixing with another LO signal (triggered by the same 25MHz clock signal mentioned above) of 
frequency 1097.8MHz. A bandpass filter with center frequency 191.4MHz and bandwidth 
${\approx}$6MHz is used at the output of the `mixer'. In the analog receiver, the 191.4MHz signal is further amplified and downconverted to 10.7MHz by mixing with a LO signal (triggered by the same 25MHz clock signal mentioned above) of frequency 180.7MHz. Here a bandpass filter with center frequency 10.7 MHz and 
bandwidth ${\approx}$1MHz is used at the output of the `mixer'. Some of the aforementioned LO and IF frequencies were chosen for the analog receiver system since the related RF components are available with us from our ongoing public outreach 
program\footnote{\textit{see}~\url{https://www.iiap.res.in/centers/radio\#pop}}.

The 10.7MHz IF signal is split into in-phase and quadrature phase components using a quadrature power splitter and then connected to a 1-bit correlator which can be assembled with simple digital logic circuits (refer Figure~\ref{figtwo}). The 10.7MHz signal from the analog receiver are digitized using a 2-level (+1 or 0) high-speed comparator (AD790). The digitized signals are then sampled in a D-type flip-flop (74LS74) at a rate of 4MHz. Later they are passed on to an Ex-NOR gate for correlation. The output of the correlator are counted with the help of a 24-bit counter, which acts as an integrator \citep{Ramesh06a}. The counter output (i.e. the correlation count) are read by a computer via a 8-bit micro-controller (Microchip's PIC16f877A). The integration time is ${\approx}$1sec. Note that due to its simple design, the sensitivity of a 1-bit digital correlator is limited. It is 
${\approx}0.64{\times}$ the sensitivity of a corresponding unquantized analog correlator (see for e.g. \citealp{Thompson04}). While this may not affect solar observations since the Sun is a `strong' source, 
observations of `faint' calibrator sources could be restricted. For e.g., the $T_{b}$ of a calibrator source like the Moon is ${\approx}\frac{1}{40}{\times}T_{b}$ of the Sun in the above frequency range (see Section 3).  
          
\section{Observations}
\label{S-Obs}

\begin{figure} 
\centerline{\includegraphics[width=\textwidth,clip=]{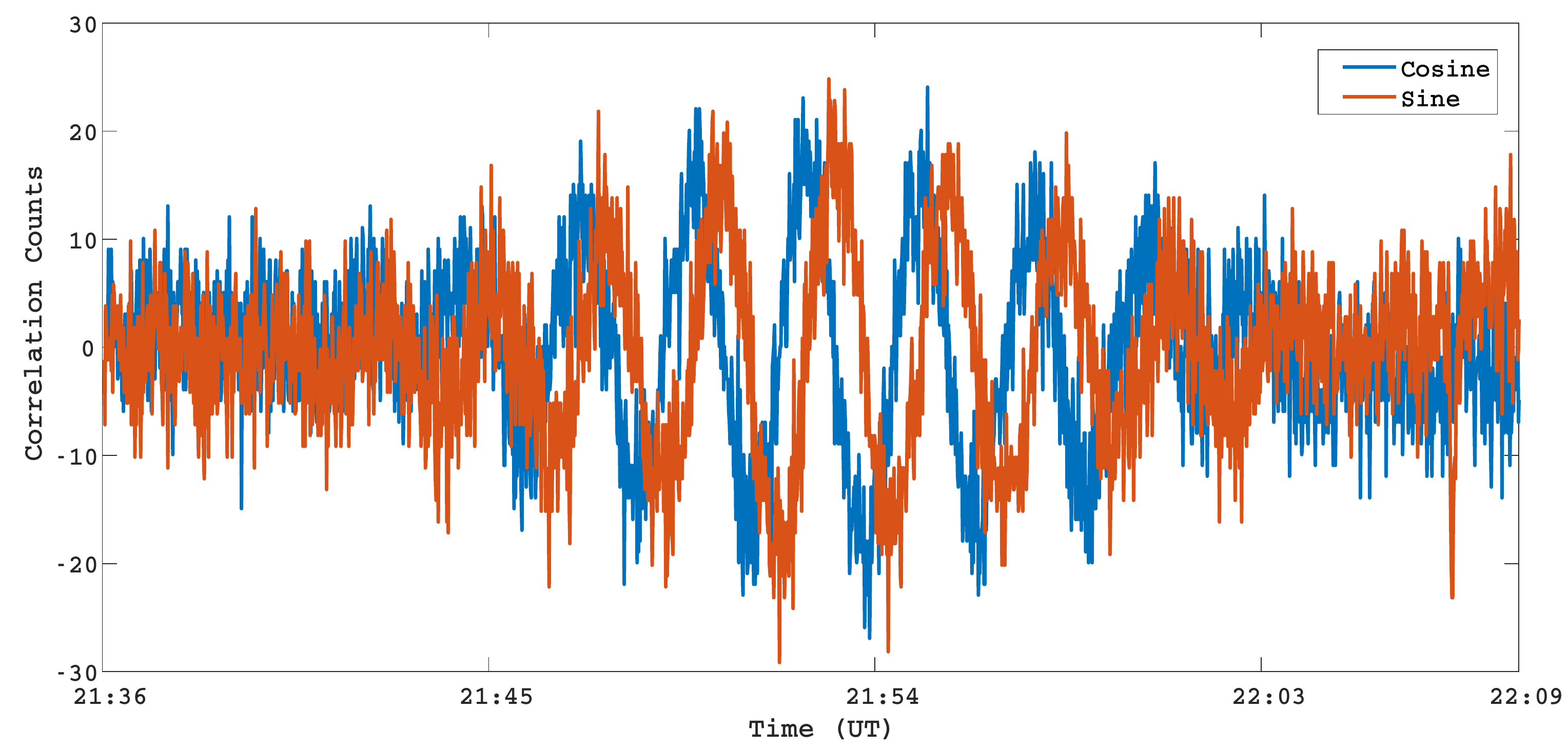}}
\caption{Observations of the Moon on 9 June 2020 at ${\approx}$11.2GHz during its transit over the local meridian in Gauribidanur.}
\label{figthree}
\end{figure}

\begin{figure} 
\centerline{\includegraphics[width=\textwidth,clip=]{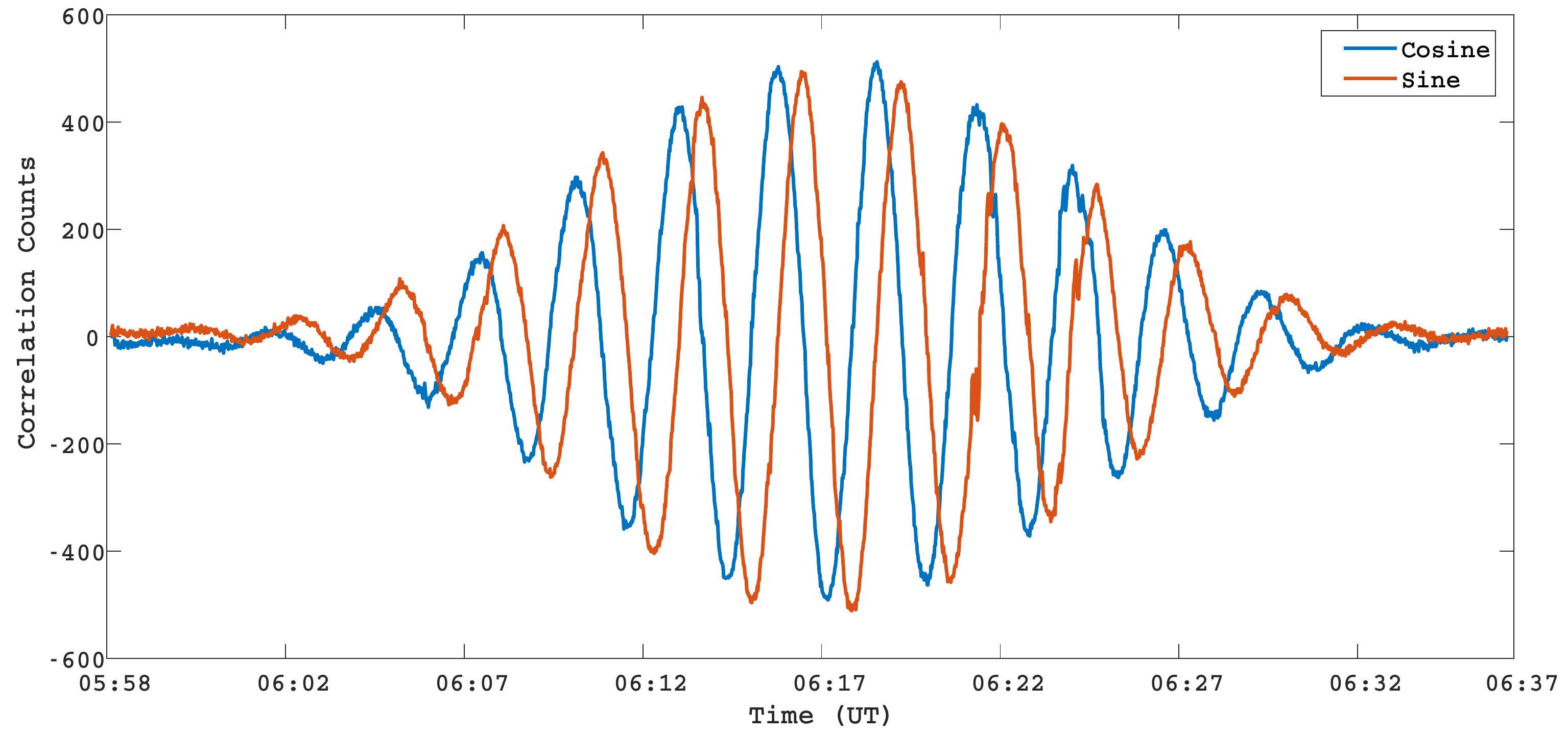}}
\caption{Same as Figure~\ref{figthree}, but observations of the Sun on 20 June 2020.}
\label{figfour}
\end{figure}

Figures~\ref{figthree} and \ref{figfour} show observations of the Moon (9 June 2020) and the Sun (20 June 2020) with the above described set-up (i.e. the Ku-band interferometer). Note that the average angular size of the Moon is ${\approx}31^{\prime}$ (see for e.g. \citealp{Kraus86}). So it can be also considered as a `point' source for our observations, similar to the Sun (see Section 2). The observations were carried out in the drift scan mode after tilting the two dish antennas towards the direction of the sources using the motors as mentioned earlier. While the `cosine' fringes correspond to the correlation between in-phase 10.7MHz IF signal from the two antennas, the `sine' fringes correspond to the correlation between in-phase 10.7MHz IF signal from one of the antennas and quadrature phase 10.7MHz IF signal from the other antenna. The observed fringe spacing  
in Figure~\ref{figfour} is 
${\approx}40^{\prime}$. The half-width of the fringe envelope is 
${\approx}3.2^{\circ}$. The observations were carried out when hour angle of the Sun was ${\approx}9^{\circ}$ to the east of 
the local meridian in Gauribidanur. The declination of the Sun was 
${\approx}23^{\circ}30^{\prime}$N. Taking into consideration the projected baseline length (${\approx}$2.3m) as seen by the source, the aforementioned fringe spacing derived from the observations is reasonably consistent with the expected value mentioned earlier. The extent of the observed fringe envelope is supposed to be same as the theoretical HPBW (${\approx}3.02^{\circ}$) of the individual dish antenna. But it is ${\approx}1.06{\times}$ wider. In fact estimates from observations on different days also gave the same result (i.e. ${\approx}3.2^{\circ}$). Therefore it is likely  that the above difference is most likely due to the design parameters of the dish. For e.g. if the electric field distribution across the aperture is parabolic, then HPBW${\approx}73^{\circ}\lambda$/D (see for e.g. \citealp{Kraus86}).
 
Maximum radiation from the Moon is usually received at Earth ${\approx}$3$-$4 days after a Full Moon day\footnote{\textit{see}~\url{https://doi.org/10.3929/ethz-a-004322130}}. In the present case, 5 June 2020 was a Full Moon day. Based on this, we calculated the theoretical $T_{b}$ of the Moon corresponding to the observations in 
Figure~\ref{figthree} on 9 June 2020 to be ${\approx}$227K \citep{Linsky73}. Using this, we then estimated the $T_{b}$ of the Sun in Figure~\ref{figfour} as  
$T_{b{\rm Sun}}{=}(x/y)T_{b{\rm Moon}}$, where $x$ and $y$ are the peak visibility amplitudes of the Sun and Moon, respectively. The visibility amplitudes were calculated as 
$\sqrt{C_{t}^{2}{+}S_{t}^{2}}$, where $C_{t}$ and $S_{t}$ are the amplitudes of the cosine and sine fringes at time $t$.
The resultant $T_{b}$ of the Sun was found to be ${\approx}$9266K. This is consistent with microwave brightness temperature spectrum of the `quiet' Sun \citep{Zirin91}. The possible causes of error in the $T_{b}$ of the Sun mentioned above are  uncertainities in the $T_{b}$ of the Moon \citep{Iwai17}, rainfall, and variation in sky-noise temperature ($T_{sky}$) as a function of elevation angle of the radio source.
But there were no rain during the present observations. Further, both the Sun and Moon were observed nearly at the same elevation (${\approx}4^{\circ}$ and 
${\approx}{-}5^{\circ}$, respectively). However, observations of the Sun and Moon were at different epochs and the receivers were not temperature controlled also. So there could be gain variations. We verified and corrected this by monitoring the Ku-band transmission from geostationary satellites INSAT 3A \& 4A (located at 
${\approx}93.5^{\circ}$E longitude and ${\approx}83^{\circ}$E longitude, respectively)\footnote{\textit{see}~\url{https://en.wikipedia.org/wiki/Indian_National_Satellite_System}}.
The equivalent isotropically radiated power (EIRP) of the above two satellites in the frequency range 11-11.25GHz and 11.55-11.70GHz are ${\approx}$48dBW and ${\approx}$52dBW, respectively. Note that the coordinates of Gauribidanur observatory are ${\approx}77^{\circ}26^{\prime}$E longitude and ${\approx}13^{\circ}36^{\prime}$N latitude. 
We observe the aforementioned satellite signal transmission everyday, particularly before and after the observations of the Sun and Moon. The maximum variation in a day is ${\approx}$2.8\%.
Having identified the different possible sources of error, we calculated the overall error in the $T_{b}$ of the Sun as follows:
by using 2nd order Gaussian fit to the observations of the Moon in Figure~\ref{figthree} we estimated the SNR to be 
${\approx}$4.4. Since $T_{b}$ of the Moon is ${\approx}$227K, the above SNR gives a temperature of ${\approx}$52K for the noise fluctuations in the observations. This is the system temperature ($T_{\rm{sys}}$). It depends primarily on the sky noise temperature ($T_{\rm{sky}}$) and noise temperature of the LNB (i.e. $T_{\rm{rcvr}}$). Assuming $T_{\rm{sky}}$ at 11.2GHz for observations near the zenith to be 
${\approx}$10K (see for e.g. \citealp{Kraus86}), we find $T_{\rm{rcvr}}{\approx}$42K. This is reasonably consistent with the noise temperature of the LNB mentioned in Section 2. Therefore in the present case, $T_{b}$ of the Moon is 
${\approx}$227${\pm}$52K,
and $T_{b}$ of the Sun is ${\approx}$9266${\pm}$2108K.

\subsection{The Solar Eclipse of 21 June 2020} 
\label{S-Eclipse}

\begin{figure} 
\centerline{\includegraphics[width=\textwidth,clip=]{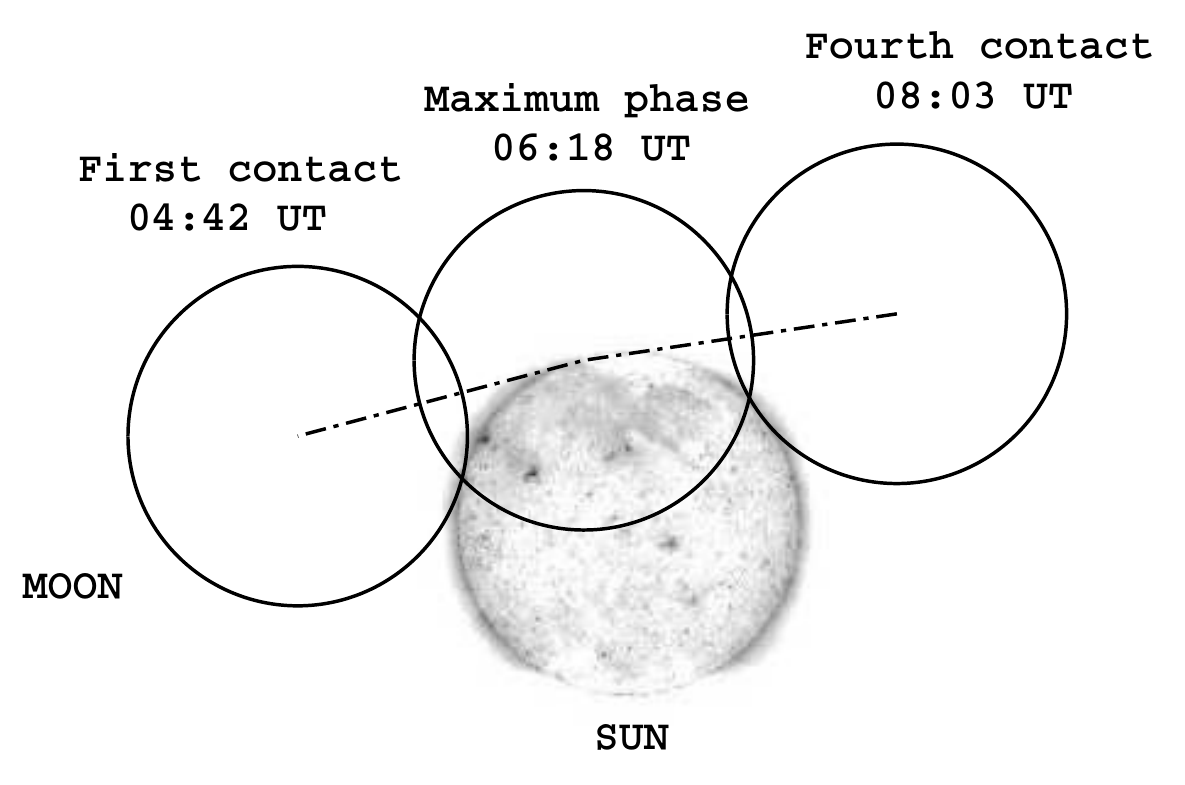}}
\caption{Positions of the Moon during the first contact, maximum phase, and fourth contact of the 21 June 2020 solar eclipse (as seen from Gauribidanur observatory) overlaid on the 211{\AA} image of the Sun obtained the same day at ${\approx}$06:30UT with the Atmospheric Imaging Assembly (AIA; \citealp{Lemen12}) on board the Solar Dynamics Observatory (SDO). North is straight up and east is to the left.}
\label{figfive}
\end{figure}

The solar eclipse of 21 June 2020 was partial at Gauribidanur observatory 
with magnitude 
${\approx}$0.493 and obscuration ${\approx}$38.33\%\footnote{\textit{see}~\url{https://www.timeanddate.com/eclipse/map/2020-june-21}}. Note that the solar eclipses are always partial at radio frequencies since the angular sizes of the `radio' Sun in the frequency range over which it is typically observed from the ground are larger compared to that of the Moon. While the average angular size of the Moon is 
${\approx}31^{\prime}$ (see Section 3), the corresponding values for the `radio' Sun at 11GHz and 80MHz (the frequencies mentioned in this paper) are 
${\approx}33^{\prime}$
and ${\approx}38^{\prime}$, respectively \citep{Borovik80,Ramesh06b}. The first contact of Moon with the limb of the Sun (i.e. the solar photosphere) occurred at 
${\approx}$04:42UT. The maximum phase of the eclipse was at ${\approx}$06:18UT. The fourth contact was at 
${\approx}$08:03UT (Figure~\ref{figfive}). Based on our experience with radio observations at low frequencies during solar eclipses in the past \citep{Ramesh99,Ramesh12a,Kathiravan11}, we carried out observations of the Sun with the Ku-band interferometer at $\approx$11.2GHz for a few days around 21 June 2020 in the drift scan mode as mentioned earlier. 

\begin{figure} 
\centerline{\includegraphics[width=\textwidth,clip=]{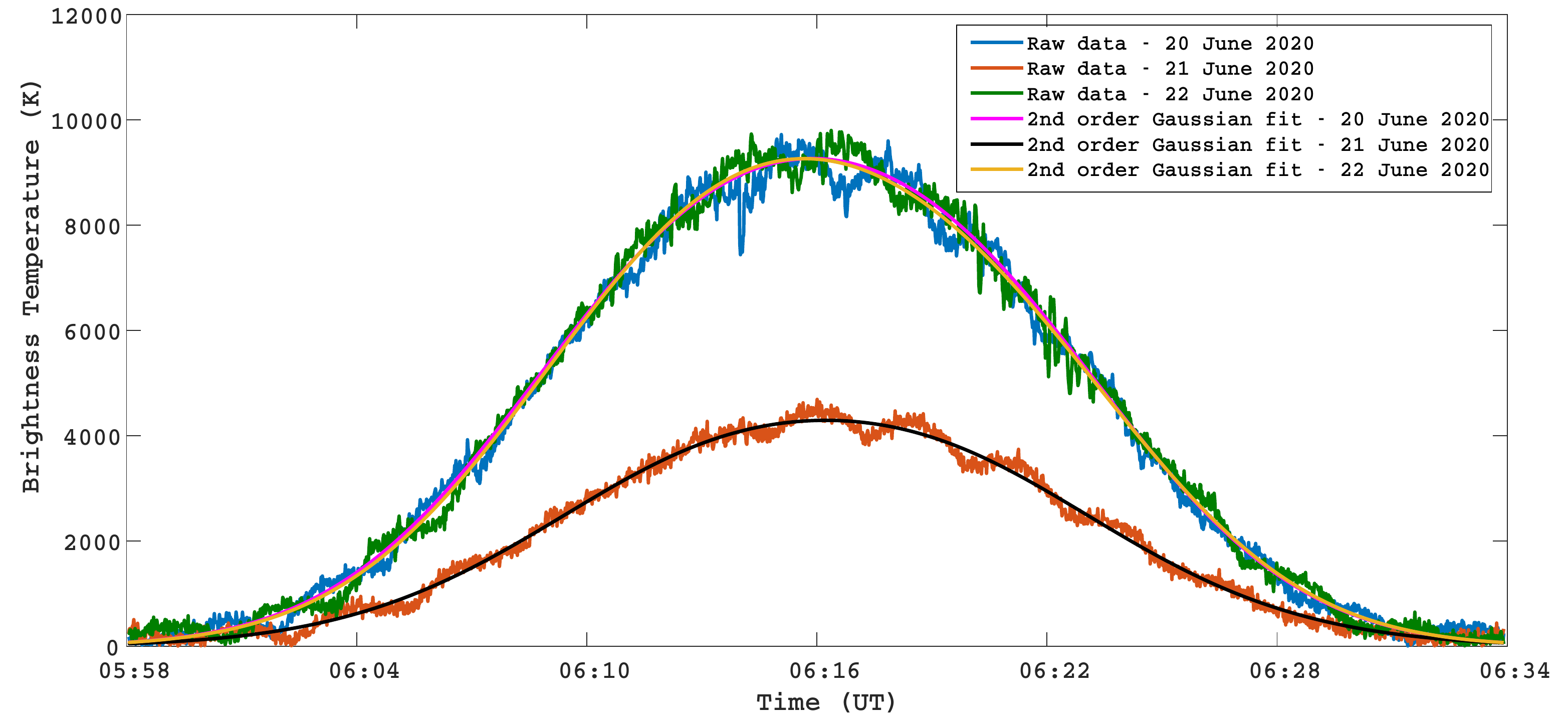}}
\caption{Observations of the Sun with Ku-band interferometer on 20, 21, \& 22 June 2020 in the transit mode after tilting the antennas towards the direction of the Sun.}
\label{figsix}
\end{figure}

Figure~\ref{figsix} shows the results of our observations on 20 June 2020 (the day before the eclipse), 21 June 2020 (during the eclipse), and 22 June 2020 (day after the eclipse). The observations on 21 June 2020 were during the maximum phase of the eclipse. The profiles shown correspond to the visibility amplitudes calculated from the observed cosine and sine fringes 
as described in Section 3.
The estimated $T_{b}$ of the Sun at $\approx$11.2GHz (after calibration using observations of the Moon with the same set-up) on the above three days are ${\approx}$9266K, 4294K, and 9263K, respectively. The $T_{b}$ 
on 21 June 2020 is lesser by ${\approx}$54\%. 
A closer look of Figure~\ref{figfive} indicates that three noticeable active regions in the northern hemisphere of the Sun were fully occulted during the maximum phase of the eclipse. This could be the reason for the aforementioned reduction in $T_{b}$ (though the eclipse obscuration was only 38\%) since radio emission associated with the active regions constitute a significant fraction of the total emission from the Sun (see for e.g. \citealp{Covington47,Mayfield71,Kathiravan11}).
An inspection of data obtained with the Gauribidanur RAdioheliograPH (GRAPH; \citealp{Ramesh98}, \citealp{Ramesh14}) at 53MHz and 80MHz during the eclipse on 21 June 2020 also reveal a similar reduction in the estimated $T_{b}$ as compared to the corresponding values observed on the day before as well as after the eclipse. 

\section{Summary} 
\label{S-Sum}

We have reported successful radio interferometric observations of the Sun at 
${\approx}$11.2GHz using two commercial dish TV antennas operating in the Ku-band 
(${\approx}$10.7-11.7GHz). The results obtained are encouraging for our plans to set up an array of such antennas for dedicated and synoptic two-dimensional spectroscopic imaging observations of the Sun in the above frequency range with minimal budget. Work is in progress to fine tune the pointing accuracy of the dishes and develop a FPGA based spectrocorrelator to observe over the entire 10.7-11.7 GHz frequency band. 

\begin{acks}
We express our gratitude to the staff of the Gauribidanur observatory for their help in setting up the antenna/receiver systems, and carrying out the observations. The SDO/AIA data are courtesy of the NASA/SDO and the AIA science teams. We thank the referee for his/her kind comments which helped us to improve the manuscript. \\
\\
{\bf Declaration:} The authors declare that there are no conflicts of interest.
\end{acks}



\begin{thebibliography}{}
  \bibitem[\protect\citeauthoryear{{Berger}}{2003}]{Berger03b}
Berger,~M.A.: 
2003, in Ferriz-Mas, A., N{\'u}{\~n}ez, M. (eds.),
    \textit{Advances in Nonlinear Dynamics}, Taylor and Francis Group, 
    London, 345.
  \bibitem[\protect\citeauthoryear{{Berger} and {Field}}{1984}]{BergerF84b}
Berger,~M.A., Field,~G.B.: 
1984, \textit{J. Fluid. Mech.} \textbf{147}, 133.
  \bibitem[\protect\citeauthoryear{{Brown}, {Canfield}, and
                                   {Pevtsov}}{1999}]{Brown99b}
Brown,~M., Canfield,~R., Pevtsov,~A.:
1999, Magnetic Helicity in Space and Laboratory Plasmas, Geophys. Mon. 
      Ser. 111, AGU.
 \bibitem[\protect\citeauthoryear{{Dupont}, {Schmidt}, and {Koutny}}{2007}]{Dupont07}
Dupont, J.-C., Schmidt, F., Koutny, P.: 2007, \solphys{} \textbf{323}, 965. 
\end{thebibliography}

\begin{thebibliography}{41}
\ifx\bisbn     \undefined \def\bisbn  #1{ISBN #1}\fi
\ifx\binits    \undefined \def\binits#1{#1}\fi
\ifx\bauthor   \undefined \def\bauthor#1{#1}\fi
\ifx\batitle   \undefined \def\batitle#1{#1}\fi
\ifx\bjtitle   \undefined \def\bjtitle#1{\textit{#1}}\fi
\ifx\bvolume   \undefined \def\bvolume#1{\textbf{#1}}\fi
\ifx\byear     \undefined \def\byear#1{#1}\fi
\ifx\bissue    \undefined \def\bissue#1{#1}\fi
\ifx\bfpage    \undefined \def\bfpage#1{#1}\fi
\ifx\blpage    \undefined \def\blpage #1{#1}\fi
\ifx\burl      \undefined \def\burl#1{\textsf{#1}}\fi
\ifx\href      \undefined \def\href#1#2{\textsf{#2}}\fi
\ifx\betal     \undefined \def\betal{\textit{et al.}}\fi
\ifx\bctitle   \undefined \def\bctitle#1{#1}\fi
\ifx\beditor   \undefined \def\beditor#1{#1}\fi
\ifx\bbtitle   \undefined \def\bbtitle#1{\textit{#1}}\fi
\ifx\bedition  \undefined \def\bedition#1{#1}\fi
\ifx\bseriesno \undefined \def\bseriesno#1{\textbf{#1}}\fi
\ifx\blocation \undefined \def\blocation#1{#1}\fi
\ifx\bsertitle \undefined \def\bsertitle#1{\textit{#1}}\fi
\ifx\bsnm      \undefined \def\bsnm#1{#1}\fi
\ifx\bsuffix   \undefined \def\bsuffix#1{#1}\fi
\ifx\bparticle \undefined \def\bparticle#1{#1}\fi
\ifx\barticle  \undefined \def\barticle#1{}\fi
\ifx\binstitute  \undefined \def\binstitute#1{#1}\fi
\ifx\bpublisher  \undefined \def\bpublisher#1{#1}\fi
\ifx\doiurl    \undefined
  \def\doiurl#1{\href{http://dx.doi.org/#1}{\textsf{DOI}}}\fi
\ifx\arxivurl  \undefined
  \def\arxivurl#1{\href{http://arxiv.org/abs/#1}{\textsf{arXiv}}}\fi
\ifx\adsurl    \undefined
  \def\adsurl#1{\href{http://adsabs.harvard.edu/abs/#1}{\textsf{ADS}}}\fi
\ifx\botherref \undefined \def\botherref#1{}\fi
\ifx\url       \undefined \def\url#1{\textsf{#1}}\fi
\ifx\bchapter  \undefined \def\bchapter#1{}\fi
\ifx\bbook     \undefined \def\bbook#1{}\fi
\ifx\bcomment  \undefined \def\bcomment#1{#1}\fi
\ifx\oauthor   \undefined \def\oauthor#1{#1}\fi
\ifx\citeauthoryear \undefined\def \citeauthoryear#1{#1}\fi
\ifx\endbibitem\undefined \def\endbibitem{}\fi
\ifx\bconflocation  \undefined \def\bconflocation#1{#1} \fi

\bibitem[\protect\citeauthoryear{{Benz} \textit{et~al.}}{2009}]{Benz09}
\begin{barticle}
\bauthor{\bsnm{{Benz}}, \binits{A.O.}},
\bauthor{\bsnm{{Monstein}}, \binits{C.}},
\bauthor{\bsnm{{Meyer}}, \binits{H.}},
\bauthor{\bsnm{{Manoharan}}, \binits{P.K.}},
\bauthor{\bsnm{{Ramesh}}, \binits{R.}},
\bauthor{\bsnm{{Altyntsev}}, \binits{A.}},
\bauthor{\bsnm{{Lara}}, \binits{A.}},
\bauthor{\bsnm{{Paez}}, \binits{J.}},
\bauthor{\bsnm{{Cho}}, \binits{K.-S.}}:
\byear{2009},
\batitle{{A World-Wide Net of Solar Radio Spectrometers: e-CALLISTO}}.
\bjtitle{Earth, Moon \& Planets}
\bvolume{104},
\bfpage{277}.
\doiurl{10.1007/s11038-008-9267-6}.
\adsurl{https://ui.adsabs.harvard.edu/abs/2009EM\%26P..104..277B/abstract}.
\end{barticle}
\endbibitem

\bibitem[\protect\citeauthoryear{{Borovik}}{1980}]{Borovik80}
\begin{barticle}
\bauthor{\bsnm{{Borovik}}, \binits{V.N.}}:
\byear{1980},
\batitle{{RATAN-600 Measurements of the 2cm-4cm Radio Brightness Distribution
  Over the Disk of the Quiet Sun}}.
\bjtitle{Sov. Astron. Lett.}
\bvolume{6},
\bfpage{236}.
\doiurl{10.1007/BF00154299}.
\adsurl{https://ui.adsabs.harvard.edu/abs/1980SvAL....6..236B/abstract}.
\end{barticle}
\endbibitem

\bibitem[\protect\citeauthoryear{{Covington}}{1947}]{Covington47}
\begin{barticle}
\bauthor{\bsnm{{Covington}}, \binits{A.E.}}:
\byear{1947},
\batitle{{MicroWave Solar Noise Observations During the Partial Eclipse of
  November 23, 1946}}.
\bjtitle{Nature}
\bvolume{159},
\bfpage{405}.
\doiurl{10.1038/159405a0}.
\adsurl{https://ui.adsabs.harvard.edu/abs/1947Natur.159..405C/abstract}.
\end{barticle}
\endbibitem

\bibitem[\protect\citeauthoryear{{F{\"u}rst}, {Hirth}, and
  {Lantos}}{1979}]{Furst79}
\begin{barticle}
\bauthor{\bsnm{{F{\"u}rst}}, \binits{E.}},
\bauthor{\bsnm{{Hirth}}, \binits{W.}},
\bauthor{\bsnm{{Lantos}}, \binits{P.}}:
\byear{1979},
\batitle{{The radius of the Sun at centimeter waves and the brightness
  distribution across the disk}}.
\bjtitle{\solphys}
\bvolume{63},
\bfpage{253}.
\doiurl{10.1007/BF00174532}.
\adsurl{https://ui.adsabs.harvard.edu/abs/1979SoPh...63..257F/abstract}.
\end{barticle}
\endbibitem

\bibitem[\protect\citeauthoryear{{Grechnev} \textit{et~al.}}{2003}]{Grechnev03}
\begin{barticle}
\bauthor{\bsnm{{Grechnev}}, \binits{V.V.}},
\bauthor{\bsnm{{Lesovoi}}, \binits{S.V.}},
\bauthor{\bsnm{{Smolkov}}, \binits{G.Y.}},
\bauthor{\bsnm{{Krissinel}}, \binits{G.B.}},
\bauthor{\bsnm{{Zandanov}}, \binits{V.G.}}, \betal:
\byear{2003},
\batitle{{The Siberian Solar Radio Telescope: the current state of the
  instrument, observations, and data}}.
\bjtitle{\solphys}
\bvolume{216},
\bfpage{239}.
\doiurl{10.1023/A:1026153410061}.
\adsurl{https://ui.adsabs.harvard.edu/abs/2003SoPh..216..239G/abstract}.
\end{barticle}
\endbibitem

\bibitem[\protect\citeauthoryear{{Iwai} \textit{et~al.}}{2017}]{Iwai17}
\begin{barticle}
\bauthor{\bsnm{{Iwai}}, \binits{K.}},
\bauthor{\bsnm{{Shimojo}}, \binits{M.}},
\bauthor{\bsnm{{Asayama}}, \binits{S.}},
\bauthor{\bsnm{{Minamidani}}, \binits{T.}},
\bauthor{\bsnm{{White}}, \binits{S.M.}},
\bauthor{\bsnm{{Bastian}}, \binits{T.S.}},
\bauthor{\bsnm{{Saito}}, \binits{M.}}:
\byear{2017},
\batitle{{The Brightness Temperature of the Quiet Solar Chromosphere at 2.6
  mm}}.
\bjtitle{\solphys}
\bvolume{292},
\bfpage{22}.
\doiurl{10.1007/s11207-016-1044-5}.
\adsurl{https://ui.adsabs.harvard.edu/abs/2017SoPh..292...22I/abstract}.
\end{barticle}
\endbibitem

\bibitem[\protect\citeauthoryear{{Kathiravan}, {Ramesh}, and
  {Nataraj}}{2007}]{Kathiravan07}
\begin{barticle}
\bauthor{\bsnm{{Kathiravan}}, \binits{C.}},
\bauthor{\bsnm{{Ramesh}}, \binits{R.}},
\bauthor{\bsnm{{Nataraj}}, \binits{H.S.}}:
\byear{2007},
\batitle{{The Post-Coronal Mass Ejection Solar Atmosphere and Radio Noise Storm
  Activity}}.
\bjtitle{\apjl}
\bvolume{656},
\bfpage{L37}.
\doiurl{10.1086/512013}.
\adsurl{https://ui.adsabs.harvard.edu/abs/2007ApJ...656L..37K/abstract}.
\end{barticle}
\endbibitem

\bibitem[\protect\citeauthoryear{{Kathiravan}, {Ramesh}, and
  {Subramanian}}{2002}]{Kathiravan02}
\begin{barticle}
\bauthor{\bsnm{{Kathiravan}}, \binits{C.}},
\bauthor{\bsnm{{Ramesh}}, \binits{R.}},
\bauthor{\bsnm{{Subramanian}}, \binits{K.R.}}:
\byear{2002},
\batitle{{Metric Radio Observations and Ray-tracing Analysis of the Onset Phase
  of a Solar Eruptive Event}}.
\bjtitle{\apjl}
\bvolume{567},
\bfpage{L93}.
\doiurl{10.1086/339801}.
\adsurl{https://ui.adsabs.harvard.edu/abs/2002ApJ...567L..93K/abstract}.
\end{barticle}
\endbibitem

\bibitem[\protect\citeauthoryear{{Kathiravan}
  \textit{et~al.}}{2011}]{Kathiravan11}
\begin{barticle}
\bauthor{\bsnm{{Kathiravan}}, \binits{C.}},
\bauthor{\bsnm{{Ramesh}}, \binits{R.}},
\bauthor{\bsnm{{Indrajit V. Barve}}},
\bauthor{\bsnm{{Rajalingam}}, \binits{M.}}:
\byear{2011},
\batitle{{Radio Observations of the Solar Corona During an Eclipse}}.
\bjtitle{\apj}
\bvolume{730},
\bfpage{91}.
\doiurl{10.1088/0004-637X/730/2/91}.
\adsurl{https://ui.adsabs.harvard.edu/abs/2011ApJ...730...91K/abstract}.
\end{barticle}
\endbibitem

\bibitem[\protect\citeauthoryear{{Kraus}}{1986}]{Kraus86}
\begin{bbook}
\bauthor{\bsnm{{Kraus}}, \binits{J.D.}}:
\byear{1986},
\bbtitle{{Radio Astronomy 2nd Ed.}},
\bpublisher{Cygnus-Quasar},
\blocation{Ohio}.
\end{bbook}
\endbibitem

\bibitem[\protect\citeauthoryear{{Kr\"{u}ger} \textit{et~al.}}{1986}]{Kruger86}
\begin{barticle}
\bauthor{\bsnm{{Kr\"{u}ger}}, \binits{A.}},
\bauthor{\bsnm{{Hildebrandt}}, \binits{J.}},
\bauthor{\bsnm{{Bogod}}, \binits{V.M.}},
\bauthor{\bsnm{{Korzhavin}}, \binits{A.M.}},
\bauthor{\bsnm{{Ahmedov}}, \binits{S.B.}},
\bauthor{\bsnm{{Gelfreikh}}, \binits{G.B.}}:
\byear{1986},
\batitle{{A Study of RATAN-600 Observations of Solar S-Component Sources}}.
\bjtitle{\solphys}
\bvolume{105},
\bfpage{111}.
\doiurl{10.1007/BF00156381}.
\adsurl{https://ui.adsabs.harvard.edu/abs/1986SoPh..105..111K/abstract}.
\end{barticle}
\endbibitem

\bibitem[\protect\citeauthoryear{{Kumari}, {Morosan}, and
  {Kilpua}}{2021}]{Anshu21}
\begin{barticle}
\bauthor{\bsnm{{Kumari}}, \binits{A.}},
\bauthor{\bsnm{{Morosan}}, \binits{D.E.}},
\bauthor{\bsnm{{Kilpua}}, \binits{E.K.J.}}:
\byear{2021},
\batitle{{On the Occurrence of Type IV Solar Radio Bursts in Solar Cycle 24 and
  Their Association with Coronal Mass Ejections}}.
\bjtitle{\apj}
\bvolume{906},
\bfpage{79}.
\doiurl{10.3847/1538-4357/abc878}.
\adsurl{https://ui.adsabs.harvard.edu/abs/2021ApJ...906...79K/abstract}.
\end{barticle}
\endbibitem

\bibitem[\protect\citeauthoryear{{Lantos} and {Avignon}}{1975}]{Lantos75}
\begin{barticle}
\bauthor{\bsnm{{Lantos}}, \binits{P.}},
\bauthor{\bsnm{{Avignon}}, \binits{Y.}}:
\byear{1975},
\batitle{{The metric quiet Sun during two cycles of activity and the nature of
  the coronal holes}}.
\bjtitle{\aap}
\bvolume{41},
\bfpage{137}.
\adsurl{https://ui.adsabs.harvard.edu/abs/1975A\%26A....41..137L/abstract}.
\end{barticle}
\endbibitem

\bibitem[\protect\citeauthoryear{{Leblanc} and {Le Squeren}}{1969}]{Leblanc69}
\begin{barticle}
\bauthor{\bsnm{{Leblanc}}, \binits{Y.}},
\bauthor{\bsnm{{Le Squeren}}, \binits{A.M.}}:
\byear{1969},
\batitle{{Dimensions, Temperature and Electron Density of the Quiet Corona}}.
\bjtitle{\aap}
\bvolume{1},
\bfpage{239}.
\adsurl{https://ui.adsabs.harvard.edu/abs/1969A\%26A.....1..239L/abstract}.
\end{barticle}
\endbibitem

\bibitem[\protect\citeauthoryear{{Lemen} \textit{et~al.}}{2012}]{Lemen12}
\begin{barticle}
\bauthor{\bsnm{{Lemen}}, \binits{J.R.}},
\bauthor{\bsnm{{Title}}, \binits{A.M.}},
\bauthor{\bsnm{{Akin}}, \binits{D.J.}},
\bauthor{\bsnm{{Boerner}}, \binits{P.F.}},
\bauthor{\bsnm{{Chou}}, \binits{C.}}, \betal:
\byear{2012},
\batitle{{The Atmospheric Imaging Assembly (AIA) on the Solar Dynamics
  Observatory (SDO)}}.
\bjtitle{\solphys}
\bvolume{272},
\bfpage{17}.
\doiurl{10.1007/s11207-011-9776-8}.
\adsurl{https://ui.adsabs.harvard.edu/abs/2012SoPh..275...17L/abstract}.
\end{barticle}
\endbibitem

\bibitem[\protect\citeauthoryear{{Linsky}}{1973}]{Linsky73}
\begin{barticle}
\bauthor{\bsnm{{Linsky}}, \binits{J.L.}}:
\byear{1973},
\batitle{{A Recalibration of the Quiet Sun Millimeter Spectrum Based on the
  Moon as an Absolute Radiometric Standard}}.
\bjtitle{\solphys}
\bvolume{28},
\bfpage{409}.
\doiurl{10.1007/BF00152312}.
\adsurl{https://ui.adsabs.harvard.edu/abs/1973SoPh...28..409L/abstract}.
\end{barticle}
\endbibitem

\bibitem[\protect\citeauthoryear{{Mayfield}, {Chapman}, and
  {Straka}}{1971}]{Mayfield71}
\begin{barticle}
\bauthor{\bsnm{{Mayfield}}, \binits{E.B.}},
\bauthor{\bsnm{{Chapman}}, \binits{G.A.}},
\bauthor{\bsnm{{Straka}}, \binits{R.M.}}:
\byear{1971},
\batitle{{Eclipse of Radio Emission on 7 March, 1970 at 10 cm Wavelength from
  the Active Region Associated with McMath Plage 10618}}.
\bjtitle{\solphys}
\bvolume{21},
\bfpage{460}.
\doiurl{10.1007/BF00154299}.
\adsurl{https://ui.adsabs.harvard.edu/abs/1971SoPh...21..460M/abstract}.
\end{barticle}
\endbibitem

\bibitem[\protect\citeauthoryear{{Mercier} and {Chambe}}{2015}]{Mercier15}
\begin{barticle}
\bauthor{\bsnm{{Mercier}}, \binits{C.}},
\bauthor{\bsnm{{Chambe}}, \binits{G.}}:
\byear{2015},
\batitle{{Electron density and temperature in the solar corona from
  multifrequency radio imaging}}.
\bjtitle{\aap}
\bvolume{583},
\bfpage{A101}.
\doiurl{10.1051/0004-6361/201425540}.
\adsurl{https://ui.adsabs.harvard.edu/abs/2015A\%26A...583A.101M/abstract}.
\end{barticle}
\endbibitem

\bibitem[\protect\citeauthoryear{{Nakajima} \textit{et~al.}}{1994}]{Nakajima94}
\begin{barticle}
\bauthor{\bsnm{{Nakajima}}, \binits{H.}},
\bauthor{\bsnm{{Nishio}}, \binits{M.}},
\bauthor{\bsnm{{Enome}}, \binits{S.}},
\bauthor{\bsnm{{Shibasaki}}, \binits{K.}},
\bauthor{\bsnm{{Takano}}, \binits{T.}}, \betal:
\byear{1994},
\batitle{{The Nobeyama radioheliograph}}.
\bjtitle{Proc. IEEE}
\bvolume{82},
\bfpage{705}.
\adsurl{https://ui.adsabs.harvard.edu/abs/1994IEEEP..82..705N/abstract}.
\end{barticle}
\endbibitem

\bibitem[\protect\citeauthoryear{{Nita} \textit{et~al.}}{2016}]{Nita16}
\begin{barticle}
\bauthor{\bsnm{{Nita}}, \binits{G.M.}},
\bauthor{\bsnm{{Hickish}}, \binits{J.}},
\bauthor{\bsnm{{MacMahon}}, \binits{D.}},
\bauthor{\bsnm{{Gary}}, \binits{D.E.}}:
\byear{2016},
\batitle{{EOVSA Implementation of a Spectral Kurtosis Correlator for Transient
  Detection and Classification}}.
\bjtitle{J. Astron. Instr.}
\bvolume{5},
\bfpage{1641009}.
\doiurl{10.1142/S2251171716410099}.
\adsurl{https://ui.adsabs.harvard.edu/abs/2016JAI.....541009N/abstract}.
\end{barticle}
\endbibitem

\bibitem[\protect\citeauthoryear{{Ramesh}}{2011}]{Ramesh11}
\begin{bchapter}
\bauthor{\bsnm{{Ramesh}}, \binits{R.}}:
\byear{2011},
\bctitle{{Low frequency solar radio astronomy at the Indian Institute of
  Astrophysics (IIA)}}.
In: \beditor{\bsnm{{Choudhuri}}, \binits{A.R.}},
\beditor{\bsnm{{Banerjee}}, \binits{D.}} (eds.)
\bbtitle{First Asia-Pacific Solar Physics Meeting},
\bpublisher{Astron. Soc. India: Conf. Ser. {\bf 2}}, \blocation{???},
\bfpage{55}.
\adsurl{https://ui.adsabs.harvard.edu/abs/2011ASInC...2...55R/abstract}.
\end{bchapter}
\endbibitem

\bibitem[\protect\citeauthoryear{{Ramesh} and {Ebenezer}}{2001}]{Ramesh01}
\begin{barticle}
\bauthor{\bsnm{{Ramesh}}, \binits{R.}},
\bauthor{\bsnm{{Ebenezer}}, \binits{E.}}:
\byear{2001},
\batitle{{ Decameter Wavelength Observations of an Absorption Burst from the
  Sun and Its Association with an X2.0/3B Flare and the Onset of a `Halo'
  Coronal Mass Ejection}}.
\bjtitle{\apjl}
\bvolume{558},
\bfpage{L141}.
\doiurl{10.1086/323498}.
\adsurl{https://ui.adsabs.harvard.edu/abs/2001ApJ...558L.141R/abstract}.
\end{barticle}
\endbibitem

\bibitem[\protect\citeauthoryear{{Ramesh} and {Shanmugha
  Sundaram}}{2000}]{Ramesh00}
\begin{barticle}
\bauthor{\bsnm{{Ramesh}}, \binits{R.}},
\bauthor{\bsnm{{Shanmugha Sundaram}}, \binits{G.A.}}:
\byear{2000},
\batitle{{Type I radio bursts and the minimum between sunspot cycles 22 \&
  23}}.
\bjtitle{\aap}
\bvolume{364},
\bfpage{873}.
\adsurl{https://ui.adsabs.harvard.edu/abs/2000A\%26A...364..873R/abstract}.
\end{barticle}
\endbibitem

\bibitem[\protect\citeauthoryear{{Ramesh}, {Subramanian}, and
  {Sastry}}{1999}]{Ramesh99}
\begin{barticle}
\bauthor{\bsnm{{Ramesh}}, \binits{R.}},
\bauthor{\bsnm{{Subramanian}}, \binits{K.R.}},
\bauthor{\bsnm{{Sastry}}, \binits{C.V.}}:
\byear{1999},
\batitle{{Eclipse Observations of Compact Sources in the Outer Solar Corona}}.
\bjtitle{\solphys}
\bvolume{185},
\bfpage{77}.
\doiurl{10.1023/A:1005149830652}.
\adsurl{https://ui.adsabs.harvard.edu/abs/1999SoPh..185...77R/abstract}.
\end{barticle}
\endbibitem

\bibitem[\protect\citeauthoryear{{Ramesh}, {Sundara Rajan}, and
  {Sastry}}{2006a}]{Ramesh06a}
\begin{barticle}
\bauthor{\bsnm{{Ramesh}}, \binits{R.}},
\bauthor{\bsnm{{Sundara Rajan}}, \binits{M.S.}},
\bauthor{\bsnm{{Sastry}}, \binits{C.V.}}:
\byear{2006}a,
\batitle{{The 1024 channel digital correlator receiver of the Gauribidanur
  radioheliograph}}.
\bjtitle{Exp. Astron.}
\bvolume{21},
\bfpage{31}.
\doiurl{10.1007/s10686-006-9065-y}.
\adsurl{https://ui.adsabs.harvard.edu/abs/2006ExA....21...31R/abstract}.
\end{barticle}
\endbibitem

\bibitem[\protect\citeauthoryear{{Ramesh} \textit{et~al.}}{1998}]{Ramesh98}
\begin{barticle}
\bauthor{\bsnm{{Ramesh}}, \binits{R.}},
\bauthor{\bsnm{{Subramanian}}, \binits{K.R.}},
\bauthor{\bsnm{{Sundara Rajan}}, \binits{M.S.}},
\bauthor{\bsnm{{Sastry}}, \binits{C.V.}}:
\byear{1998},
\batitle{{The Gauribidanur Radioheliograph}}.
\bjtitle{\solphys}
\bvolume{181},
\bfpage{439}.
\doiurl{10.1023/A:1005075003370}.
\adsurl{https://ui.adsabs.harvard.edu/abs/1998SoPh..181..439R/abstract}.
\end{barticle}
\endbibitem

\bibitem[\protect\citeauthoryear{{Ramesh} \textit{et~al.}}{2006b}]{Ramesh06b}
\begin{barticle}
\bauthor{\bsnm{{Ramesh}}, \binits{R.}},
\bauthor{\bsnm{{Nataraj}}, \binits{H.S.}},
\bauthor{\bsnm{{Kathiravan}}, \binits{C.}},
\bauthor{\bsnm{{Sastry}}, \binits{C.V.}}:
\byear{2006}b,
\batitle{{The Equatorial Background Solar Corona during Solar Minimum}}.
\bjtitle{\apj}
\bvolume{648},
\bfpage{707}.
\doiurl{10.1086/505677}.
\adsurl{https://ui.adsabs.harvard.edu/abs/2006ApJ...648..707R/abstract}.
\end{barticle}
\endbibitem

\bibitem[\protect\citeauthoryear{{Ramesh} \textit{et~al.}}{2012a}]{Ramesh12a}
\begin{barticle}
\bauthor{\bsnm{{Ramesh}}, \binits{R.}},
\bauthor{\bsnm{{Kathiravan}}, \binits{C.}},
\bauthor{\bsnm{{Indrajit V. Barve}}},
\bauthor{\bsnm{{Rajalingam}}, \binits{M.}}:
\byear{2012}a,
\batitle{{High Angular Resolution Radio Observations of a Coronal Mass Ejection
  Source Region at Low Frequencies during a Solar Eclipse}}.
\bjtitle{\apj}
\bvolume{744},
\bfpage{165}.
\doiurl{10.1088/0004-637X/744/2/165}.
\adsurl{https://ui.adsabs.harvard.edu/abs/2012ApJ...744..165R/abstract}.
\end{barticle}
\endbibitem

\bibitem[\protect\citeauthoryear{{Ramesh} \textit{et~al.}}{2012b}]{Ramesh12b}
\begin{barticle}
\bauthor{\bsnm{{Ramesh}}, \binits{R.}},
\bauthor{\bsnm{{Kathiravan}}, \binits{C.}},
\bauthor{\bsnm{{Anna Lakshmi}}, \binits{M.}},
\bauthor{\bsnm{{Gopalswamy}}, \binits{N.}},
\bauthor{\bsnm{{Umapathy}}, \binits{S.}}:
\byear{2012}b,
\batitle{{The Location of Solar Metric Type II Radio Bursts with Respect to the
  Associated Coronal Mass Ejections}}.
\bjtitle{\apj}
\bvolume{752},
\bfpage{107}.
\doiurl{10.1088/0004-637X/752/2/107}.
\adsurl{https://ui.adsabs.harvard.edu/abs/2012ApJ...752..107R/abstract}.
\end{barticle}
\endbibitem

\bibitem[\protect\citeauthoryear{{Ramesh} \textit{et~al.}}{2014}]{Ramesh14}
\begin{bchapter}
\bauthor{\bsnm{{Ramesh}}, \binits{R.}},
\bauthor{\bsnm{{Kathiravan}}, \binits{C.}},
\bauthor{\bsnm{{Sundara Rajan}}, \binits{M.S.}},
\bauthor{\bsnm{{Indrajit V. Barve}}},
\bauthor{\bsnm{{Rajalingam}}, \binits{M.}}:
\byear{2014},
\bctitle{{Solar observations at low frequencies with the Gauribidanur
  radioheliograph}}.
In: \beditor{\bsnm{{Chengalur}}, \binits{J.N.}},
\beditor{\bsnm{{Gupta}}, \binits{Y.}} (eds.)
\bbtitle{The Metrewavelength Sky},
\bpublisher{Astron. Soc. India: Conf. Ser. {\bf 13}}, \blocation{???},
\bfpage{19}.
\adsurl{https://ui.adsabs.harvard.edu/abs/2014ASInC..13...19R/abstract}.
\end{bchapter}
\endbibitem

\bibitem[\protect\citeauthoryear{{Ramesh} \textit{et~al.}}{2020}]{Ramesh20}
\begin{barticle}
\bauthor{\bsnm{{Ramesh}}, \binits{R.}},
\bauthor{\bsnm{{Kumari}}, \binits{A.}},
\bauthor{\bsnm{{Kathiravan}}, \binits{C.}},
\bauthor{\bsnm{{Ketaki}}, \binits{D.}},
\bauthor{\bsnm{{Rajesh}}, \binits{M.}},
\bauthor{\bsnm{{Vrunda}}, \binits{M.}}:
\byear{2020},
\batitle{{Low-Frequency Radio Observations of the Quiet Corona During the
  Descending Phase of Sunspot Cycle 24}}.
\bjtitle{\grl}
\bvolume{47},
\bfpage{e90426}.
\doiurl{10.1029/2020GL090426}.
\adsurl{https://ui.adsabs.harvard.edu/abs/2020GeoRL..4790426R/abstract}.
\end{barticle}
\endbibitem

\bibitem[\protect\citeauthoryear{{Saint-Hilaire}, {Vilmer}, and
  {Kerdraon}}{2013}]{Saint13}
\begin{barticle}
\bauthor{\bsnm{{Saint-Hilaire}}, \binits{P.}},
\bauthor{\bsnm{{Vilmer}}, \binits{N.}},
\bauthor{\bsnm{{Kerdraon}}, \binits{A.}}:
\byear{2013},
\batitle{{A Decade of Solar Type III Radio Bursts Observed by the Nancay
  Radioheliograph 1998-2008}}.
\bjtitle{\apj}
\bvolume{762},
\bfpage{60}.
\doiurl{10.1088/0004-637X/762/1/60}.
\adsurl{https://ui.adsabs.harvard.edu/abs/2013ApJ...762...60S/abstract}.
\end{barticle}
\endbibitem

\bibitem[\protect\citeauthoryear{{Sasikumar Raja}
  \textit{et~al.}}{2014}]{Sasikumar14}
\begin{barticle}
\bauthor{\bsnm{{Sasikumar Raja}}, \binits{K.}},
\bauthor{\bsnm{{Ramesh}}, \binits{R.}},
\bauthor{\bsnm{{Hariharan}}, \binits{K.}},
\bauthor{\bsnm{{Kathiravan}}, \binits{C.}},
\bauthor{\bsnm{{Wang}}, \binits{T.J.}}:
\byear{2014},
\batitle{{An Estimate of the Magnetic Field Strength Associated with a Solar
  Coronal Mass Ejection from Low Frequency Radio Observations}}.
\bjtitle{\apj}
\bvolume{796},
\bfpage{56}.
\doiurl{10.1088/0004-637X/796/1/56}.
\adsurl{https://ui.adsabs.harvard.edu/abs/2014ApJ...796...56S/abstract}.
\end{barticle}
\endbibitem

\bibitem[\protect\citeauthoryear{{Sastry}}{1995}]{Sastry95}
\begin{barticle}
\bauthor{\bsnm{{Sastry}}, \binits{C.V.}}:
\byear{1995},
\batitle{{The Decameter and Meter Wave Radiotelescopes in India and
  Mauritius}}.
\bjtitle{\ssr}
\bvolume{72},
\bfpage{629}.
\doiurl{10.1007/BF00749008}.
\adsurl{https://ui.adsabs.harvard.edu/abs/1995SSRv...72..629S/abstract}.
\end{barticle}
\endbibitem

\bibitem[\protect\citeauthoryear{{Selhorst} \textit{et~al.}}{2011}]{Selhorst11}
\begin{barticle}
\bauthor{\bsnm{{Selhorst}}, \binits{C.L.}},
\bauthor{\bsnm{{Gim{\'e}nez De Castro}}, \binits{C.G.}},
\bauthor{\bsnm{{V{\'a}lio}}, \binits{A.}},
\bauthor{\bsnm{{Costa}}, \binits{J.E.R.}},
\bauthor{\bsnm{{Shibasaki}}, \binits{K.}}:
\byear{2011},
\batitle{{The Behavior of the 17 GHz Solar Radius and Limb Brightening in the
  Spotless Minimum XXIII/XXIV}}.
\bjtitle{\apj}
\bvolume{734},
\bfpage{64}.
\doiurl{10.1088/0004-637X/734/1/64}.
\adsurl{https://ui.adsabs.harvard.edu/abs/2011ApJ...734...64S/abstract}.
\end{barticle}
\endbibitem

\bibitem[\protect\citeauthoryear{{Shibasaki}}{2013}]{Shibasaki13}
\begin{barticle}
\bauthor{\bsnm{{Shibasaki}}, \binits{K.}}:
\byear{2013},
\batitle{{Long-Term Global Solar Activity Observed by the Nobeyama
  Radioheliograph}}.
\bjtitle{Publ. Astron. Soc. Japan}
\bvolume{65},
\bfpage{S17}.
\doiurl{10.1093/pasj/65.sp1.S17}.
\adsurl{https://ui.adsabs.harvard.edu/abs/2013PASJ...65S..17S/abstract}.
\end{barticle}
\endbibitem

\bibitem[\protect\citeauthoryear{{Tanaka} \textit{et~al.}}{1973}]{Tanaka73}
\begin{barticle}
\bauthor{\bsnm{{Tanaka}}, \binits{H.}},
\bauthor{\bsnm{{Castelli}}, \binits{J.P.}},
\bauthor{\bsnm{{Covington}}, \binits{A.E.}},
\bauthor{\bsnm{{Kr{\"u}ger}}, \binits{A.}},
\bauthor{\bsnm{{Landecker}}, \binits{T.L.}},
\bauthor{\bsnm{{Tlamicha}}, \binits{A.}}:
\byear{1973},
\batitle{{Absolute Calibration of Solar Radio Flux Density in the Microwave
  Region}}.
\bjtitle{\solphys}
\bvolume{29},
\bfpage{243}.
\doiurl{10.1007/BF00153452}.
\adsurl{https://ui.adsabs.harvard.edu/abs/1973SoPh...29..243T/abstract}.
\end{barticle}
\endbibitem

\bibitem[\protect\citeauthoryear{{Tapping}}{2013}]{Tapping13}
\begin{barticle}
\bauthor{\bsnm{{Tapping}}, \binits{K.F.}}:
\byear{2013},
\batitle{{The 10.7 cm solar radio flux ($\rm F_{10.7}$)}}.
\bjtitle{Space Weather}
\bvolume{11},
\bfpage{394}.
\doiurl{10.1002/swe.20064}.
\adsurl{https://ui.adsabs.harvard.edu/abs/2013SpWea..11..394T/abstract}.
\end{barticle}
\endbibitem

\bibitem[\protect\citeauthoryear{{Thompson}, {Moran}, and {Swenson
  Jr.}}{2004}]{Thompson04}
\begin{bbook}
\bauthor{\bsnm{{Thompson}}, \binits{A.R.}},
\bauthor{\bsnm{{Moran}}, \binits{J.M.}},
\bauthor{\bsnm{{Swenson Jr.}}, \binits{G.W.}}:
\byear{2004},
\bbtitle{{Interferometry and Synthesis in Radio Astronomy 2nd Ed.}},
\bpublisher{WILEY-VCH},
\blocation{Weinheim}.
\end{bbook}
\endbibitem

\bibitem[\protect\citeauthoryear{{Yan} \textit{et~al.}}{2021}]{Yan21}
\begin{barticle}
\bauthor{\bsnm{{Yan}}, \binits{Y.}},
\bauthor{\bsnm{{Chen}}, \binits{Z.}},
\bauthor{\bsnm{{Wang}}, \binits{W.}},
\bauthor{\bsnm{{Liu}}, \binits{F.}},
\bauthor{\bsnm{{Geng}}, \binits{L.}},
\bauthor{\bsnm{{Chen}}, \binits{L.}},
\bauthor{\bsnm{{Tan}}, \binits{C.}},
\bauthor{\bsnm{{Chen}}, \binits{X.}},
\bauthor{\bsnm{{Su}}, \binits{C.}},
\bauthor{\bsnm{{Tan}}, \binits{B.}}:
\byear{2021},
\batitle{{Mingantu Spectral Radioheliograph for Solar and Space Weather
  Studies}}.
\bjtitle{Frontiers Astron. Space Sci.}
\bvolume{8},
\bfpage{20}.
\doiurl{10.3389/fspas.2021.584043}.
\adsurl{https://ui.adsabs.harvard.edu/abs/2021FrASS...8...20Y/abstract}.
\end{barticle}
\endbibitem

\bibitem[\protect\citeauthoryear{{Zirin}, {Baumert}, and
  {Hurford}}{1991}]{Zirin91}
\begin{barticle}
\bauthor{\bsnm{{Zirin}}, \binits{H.}},
\bauthor{\bsnm{{Baumert}}, \binits{B.M.}},
\bauthor{\bsnm{{Hurford}}, \binits{G.}}:
\byear{1991},
\batitle{{The Microwave Brightness Temperature Spectrum of the Quiet Sun}}.
\bjtitle{\apj}
\bvolume{370},
\bfpage{779}.
\doiurl{10.1086/169861}.
\adsurl{https://ui.adsabs.harvard.edu/abs/1991ApJ...370..779Z/abstract}.
\end{barticle}
\endbibitem

\end{thebibliography}

\end{article} 

\end{document}